\documentclass[a4paper]{llncs}
\usepackage{amssymb}
\usepackage{amsmath}
\usepackage{graphicx}
\usepackage{subfigure}
\usepackage{listings}
\usepackage[ruled, noend, noline]{algorithm2e}
\usepackage{stmaryrd}
\usepackage{soul}
\usepackage[bookmarks,bookmarksopen,bookmarksdepth=2]{hyperref}
\usepackage{multirow}
\usepackage{tabu}
\usepackage{rotating}
\usepackage{wrapfig}
\usepackage{todonotes}
\usepackage{fancyvrb,xcolor}

\usepackage{tikz}
\usetikzlibrary{positioning,shapes,arrows,chains}
\usetikzlibrary{matrix}
\usetikzlibrary{arrows,automata}
\usetikzlibrary{calc,arrows}
\usepackage[european]{circuitikz}

\newcommand{\linefill}{\cleaders\hbox{$\smash{\mkern-2mu\mathord-\mkern-2mu}$}\hfill\vphantom{\lower1pt\hbox{$\rightarrow$}}}  
\newcommand{\Linefill}{\cleaders\hbox{$\smash{\mkern-2mu\mathord=\mkern-2mu}$}\hfill\vphantom{\hbox{$\Rightarrow$}}}  
\newcommand{\transi}[2]{\mathrel{\lower1pt\hbox{$\mathrel-_{\vphantom{#2}}\mkern-8mu\stackrel{#1}{\linefill_{\vphantom{#2}}}\mkern-11mu\rightarrow_{#2}$}}}
\newcommand{\trans}[1]{\transi{#1}{{}}}
\newcommand{\transo}{\mathord{\trans{~}}}
\newcommand{\ntransi}[2]{\mathrel{\lower1pt\hbox{$\mathrel-_{\vphantom{#2}}\mkern-8mu\stackrel{#1}{\linefill_{\vphantom{#2}}}\mkern-8mu\nrightarrow_{#2}$}}}

\newcommand{\Transi}[2]{\mathrel{\lower1pt\hbox{$\mathrel=_{\vphantom{#2}}\mkern-8mu\stackrel{#1}{\Linefill_{\vphantom{#2}}}\mkern-11mu\Rightarrow_{#2}$}}}

\newcommand{\symTC}{\triangleright}

\begin{document}

\title{Deriving approximation tolerance constraints from verification runs\thanks{This work was partially supported by the German Research Foundation (DFG) within the Collaborative Research Centre ``On-The-Fly Computing'' (SFB 901).}}
\author{Tobias Isenberg \and Marie-Christine Jakobs \and Felix Pauck \and Heike Wehrheim}
\institute{	
   Paderborn University, 	Germany\\
	\email{\{isenberg,marie.christine.jakobs,wehrheim\}@upb.de}
}
\maketitle

\begin{abstract} Approximate computing (AC) is an emerging paradigm for energy-efficient
computation. The basic idea of AC is to sacrifice high precision for low energy
by allowing for hardware which only carries out "approximately correct" calculations. For software verification, this challenges the validity of verification results for programs run on approximate hardware. 

In this paper, we present a novel approach to examine program correctness in the context of approximate computing.
In contrast to all existing approaches, we start with a standard program verification and compute the allowed {\em tolerances} for AC hardware from that verification run. More precisely, we derive a set of constraints which -- when met by the AC hardware -- guarantees the verification result to carry over to AC. Our approach is based on the framework of {\em abstract interpretation}. On the practical side, we furthermore (1) show  how to extract tolerance constraints from verification runs employing predicate abstraction as an instance of abstract interpretation, and (2) show how to check such constraints on hardware designs. We exemplify our technique on example C programs and a number of recently proposed {\em approximate adders}. 
\end{abstract}

\section{Introduction}

Approximate computing (AC) \cite{DBLP:journals/cacm/Kugler15a,DBLP:conf/ets/HanO13}  is a new computing paradigm which aims at reducing energy consumption at the cost of computation {\em precision}. A number of application domains can tolerate AC because they are inherently resilient to imprecision (e.g., machine learning, big data analytics, image processing, speech recognition). Computation precision can be reduced by either directly manipulating program executions on the algorithmic level (e.g.\ by loop perforation \cite{DBLP:conf/pepm/CarbinKMR13}) or by employing approximate hardware for program execution \cite{DBLP:conf/pldi/SampsonDFGCG11}. Approximation on the level of hardware can be achieved by techniques like voltage overscaling  or by directly making imprecise hardware designs with less chip area. The approximate adders which we will later use employ the latter technique, and simply have limited carry propagation. 

For software verification, the use of approximate hardware challenges soundness and raises the question of whether the achieved verification result will really be valid when the program is being executed.  So far, correctness in the context of approximate computing has either studied {\em quantitative reliability}, i.e., the probability that outputs of functions have correct values \cite{DBLP:conf/oopsla/CarbinMR13,DBLP:conf/oopsla/MisailovicCAQR14} (employed for the language Rely), or differences between approximate and precise executions \cite{Lahiri2015,Gopa} (applying differential program verification). Alternatively, some approaches plainly use types and type checking to separate the program into precise and approximate parts  (language EnerJ) \cite{DBLP:conf/pldi/SampsonDFGCG11}. 
All of these techniques take a hardware-centric approach: take the (non-)guarantees of the hardware, and develop new analysis methods working under such weak guarantees. The opposite direction, namely use standard program analysis procedures and have the verification impose constraints on the allowed approximation, has not been studied so far. This is despite the fact that such an approach directly allows re-use of existing verification technology for program verification as well as for checking the constraints on the approximate hardware. Another advantage of this approach is that the imposed constraints can be checked on multiple hardware designs, as we did in our examples. 

\smallskip
\noindent In this paper, we propose a new strategy for making software verification reliable for approximate computing. Within the broad spectrum of AC techniques, we focus on {\em deterministic} approximate designs, i.e., approximate hardware designs with deterministic truth tables. 
We start with a verification run proving safety of a program.
For the moment on, we assume that the safety property is encoded with assertions or specific error labels \texttt{ERR}.
With a proper instrumentation of the program various properties can be encoded in such a way.
In Section~\ref{sec:termination} we exemplary describe how to encode termination proofs.

Our approach derives from that verification run requirements on the hardware executing the program. We call such requirements {\em tolerance constraints}. A tolerance constraint acts like a pre/postcondition pair and describes the expected output of a hardware design when supplied with specified inputs. The derived tolerance constraints capture the assumptions the verification run has made on the executing hardware. Thus, they are specific to the program and safety property under consideration. Tolerance constraints refer to program statements, e.g., statements using addition as operation. Typically, tolerance constraints are much less restrictive than the precise truth table of a hardware operation would dictate. 

To instantiate this general idea, we had to select the underlying verification technique.
We discuss the alternatives in Section~\ref{sec:discuss}, after we presented our concrete instantiation.
In the following, we formulate the derivation of tolerance constraints within the framework of {\em abstract interpretation}, thus, making the technique applicable to all abstract interpretation based program analyses. We prove soundness of our technique by showing that a program, which has been proven correct, will also run correctly on AC when the employed approximate hardware satisfies the derived tolerance constraints. 

\begin{figure}[t]
\begin{minipage}[t]{.05\textwidth}\ 
\end{minipage}
\begin{minipage}[t]{.5\textwidth}
\begin{verbatim}
  int arr[1000];

  for(int j:=0;j<990;) {
    j:=j+10;
    if (!(j>=0 && j<1000))
       ERR: ; 
    arr[j]:=0;
  } 
\end{verbatim} 
\end{minipage} \quad
\begin{minipage}[t]{.3\textwidth}
\begin{verbatim}
  int u:=input();
  int sum:=1;

  if (u>0) 
     sum:=1+u;
  if (sum==0)
    ERR: ; 
\end{verbatim}
\end{minipage}
\caption{Programs $Array$ (left) and $AddOne$ (right)}
\label{fig:array}
\end{figure}

To see our technique in practice, we instantiate the general framework based on abstract interpretation with predicate abstraction \cite{Graf1997,DBLP:journals/sttt/BallPR03}. In this case, tolerance constraints are pairs $(p,q)$ of predicates on inputs and expected outputs of a hardware operation. As a first example, take a look at the left program in Figure \ref{fig:array}. The left program writes to an array within a for-loop. The property to be checked (encoded as an error state \verb+ERR+) is an array-index-inside-bounds check. Using $x$ and $y$ as inputs and $z$ as output (i.e., $z=x+y$), the tolerance constraint on addition (+) derived from a verification run showing correctness is \[ (x \geq 0 \wedge x \leq 989 \wedge y=10 \Rightarrow z \geq 0 \wedge z \leq 999)  \] 

\noindent It states that the hardware adder should guarantee that adding 10 to a value in between 0 and 989 never brings us outside the range $[0,999]$, and thus the program never crashes with an index-out-of-bounds exception. 

Using the analysis tool \caps{CPAchecker} \cite{CPAchecker2011} for verification runs, we implemented the extraction of tolerance constraints from abstract reachability graphs constructed during verification. The constraints will be in SMT-Lib format \cite{SMTLIB}. To complete the picture, we have furthermore implemented a procedure for {\em tolerance checking} on hardware designs. This technique constructs a specific checker circuit out of a given hardware design (in Verilog) and  tolerance constraint. We have evaluated our overall approach on example C programs, e.g., taken from the software verification competition benchmark, using as AC hardware different  approximate adders from the literature (Verilog designs taken from the website accompanying \cite{ShafiqueAHH15}). 
During evaluation, we examined if a program which uses an approximate adder still terminates, adheres to a protocol, or remains memory safe. Additionally, we looked at certain properties of additions like monotonicity to capture the different behavior of precise and approximate adders.

\section{Background} We start by formally defining the syntax and semantics of programs, and by introducing the framework of abstract interpretation \cite{DBLP:conf/popl/CousotC77}. 

\paragraph{Programs.} 
For our formal framework, we assume programs to have integer variables only\footnote{For the practical evaluation we, however, allow for arbitrary C programs.}, $\mathit{Ops} = \{+,-,*,\setminus\}$  to be the set of binary operators on integers, $\mathbb{Z}$ to be the integer constants and $\mathit{Cmp} = \{<,\leq, >, \geq, = \}$ the set of comparison operators on integers. 
Programs use variables out of a set $\mathit{Var}$, and have two sorts of statements from a set $\mathit{Stm}$: (1) conditionals \verb+assume b+ ($b$ boolean condition over $\mathit{Var}$ formed using $\mathit{Ops}$ and $\mathit{Cmp}$) and (2) assignments \verb+v:=expr+, $v\in Var$, $expr$ expression over $\mathit{Var}$ formed with $\mathit{Ops}$. Formally, programs are given by control flow automata. 

\begin{definition}
 A {\em control flow automaton} (CFA) $P=(L,\ell_0,E,\mathit{Err})$ consists of a finite set of {\em locations} $L$, an {\em initial location} $\ell_0 \in L$ and a set of {\em edges} $E \subseteq L \times \mathit{Stm} \times L$ and a set of error locations $\mathit{Err} \subseteq L$. 
\end{definition} 

\noindent Note that we mark the error locations in programs with the label \verb+ERR+ (or similar). A concrete state of a program is a mapping $s: \mathit{Var} \rightarrow \mathbb{Z}$, and $\Sigma$ is the set of all states. For a state $s$, we define a {\em  state update} wrt.\ $u\in \mathit{Var}$ and $c \in \mathbb{Z}$ to be  $s[u:=c](u) = c$, $s[u:=c](v) = s(v)$ for $u \neq v$. For a state $s$ and a boolean condition $b$, we write $s \models b$ to state that $b$ is true in $s$. A configuration of a program is a pair $(s,\ell)$, $s\in \Sigma, \ell \in L$.   

The semantics of program statements is given by the following (partial) {\em next transformers} $next_{stm} : \Sigma \rightarrow \Sigma$
with 
\[ next_{stm}(s) = s' \mbox{ with } \left\{ \begin{array}{ll} s' = s & \mbox{ if } stm \equiv \verb+assume b+ \wedge s \models b \\
                                                                       s' = s[v:=s(expr)]  & \mbox{ if } stm \equiv \verb+v:=expr+   
                                                  \end{array} \right . \]

\noindent We lift $next_{stm}$ to sets of states by $next_{stm}(S) = \{next_{stm}(s) \mid s \in S \}$. Note that this lifted function is total. The next transformers together with the control flow determine the transition system of a program. 

\begin{definition}
The {\em concrete transition system} $T(P) = (Q, q_0, \transo)$ of a CFA $P=(L,\ell_0,E,Err)$ consists of 
	\begin{itemize}
	\item a set of configurations $Q = \Sigma  \times L$,
       \item an initial configuration $q_0 = (s_0,\ell_0)$ where $s_0(v) = 0$ for all $v\in Var$,
	\item a transition relation $\transo \subseteq Q \times \mathit{Stm} \times Q$ with $(s,\ell) \trans {stm} (s',\ell')$ if $(\ell,stm,\ell') \in E$ and $next_{stm}(s)=s'$.
	\end{itemize} 
\end{definition}

\noindent An error location is {\em reachable} in $T(P)$ if there is a path from $(s_0,\ell_0)$ to a configuration $(*,\ell)$ with $\ell \in \mathit{Err}$.  If no error location is reachable, we say that the transition system is {\em free of errors}.

\paragraph{Abstract interpretation.} For verifying that a program is free of errors, we use the framework of abstract interpretation (AI) \cite{DBLP:conf/popl/CousotC77}. Thus we assume that the verification run from which we derive tolerance constraints is carried out by an analysis tool employing abstract interpretation as basic verification technology. 

Instead of concrete states, instances of AI frameworks  employ abstract domains $Abs$ and execute abstract versions of the next transformers on it.  Abstract domains are equipped with an ordering $\sqsubseteq_{Abs}$, and $(Abs,\sqsubseteq_{Abs})$ has to form a complete lattice (as does $(2^{\Sigma}, \subseteq)$). To relate abstract and concrete domain, two monotonic functions are used: an abstraction function $\alpha: 2^{\Sigma} \rightarrow Abs$ and a concretisation function $\gamma: Abs \rightarrow 2^{\Sigma}$.  
The pair $(\alpha,\gamma)$ has to form a {\em Galois connection}, i.e.\ the following has to hold: $\forall S \in 2^\Sigma: S \subseteq \gamma(\alpha(S))$ and $\forall abs \in Abs: \alpha(\gamma(abs)) \sqsubseteq_{Abs} abs$. We require the least element of the lattice $(Abs,\sqsubseteq_{Abs})$ (which we denote by $a_\bot$) to be mapped onto the least element of $(2^{\Sigma}, \subseteq)$ which is the empty set $\emptyset$. 

On the abstract domain, the AI instance defines a total abstract next transformer $next_{stm}^\# : Abs \rightarrow Abs$. To be useful for verification, the abstract transformer has to faithfully reflect the behaviour of the concrete transformer. 

\begin{definition}
An abstract next transformer $next_{stm}^\# : Abs \rightarrow Abs$ is a {\em safe approximation} of the concrete next transformer if the following holds:
\[ \forall abs \in Abs, \forall stm \in \mathit{Stm}: \alpha(next_{stm}(\gamma(abs)) \sqsubseteq_{Abs} next_{stm}^\# (abs) \]
\end{definition}

\noindent Using the abstract next transformer, we can construct an abstract transition system of a program. 

\begin{definition}
The {\em abstract transition system} $T^\#(P) = (Q,q_0,\transo)$ of a program $P$ with respect to an abstract domain $Abs$ and functions $next_{stm}^\#$ consists of 
	\begin{itemize}
	\item a set of configurations $Q = Abs  \times L$,
       \item an initial configuration $q_0 = (a_0,\ell_0)$ where $a_0 = \alpha(\{s_0\})$ with $s_0(v) = 0$ for all $v\in Var$, 
	\item a transition relation $\transo \subseteq Q \times \mathit{Stm} \times Q$ with $(a,\ell) \trans {stm} (a',\ell')$ if $(\ell,stm,\ell') \in E$ and $next_{stm}^\#(a)=a'$.
	\end{itemize}
\end{definition}

\noindent An abstract configuration \((a,\ell)\) is \emph{reachable} in $T^\#(P)$ if there is a path from $q_0=(a_0,\ell_0)$ to a configuration $(a,\ell)$.
We denote the set of reachable configurations in $T^\#(P)$ by $Reach(T^\#(P))$ or simply $Reach^\#$.
An error location is reachable in $T^\#(P)$ if there is a path from $(a_0,\ell_0)$ to a configuration $(a,\ell)$ with $\ell \in \mathit{Err}$, $a \neq a_\bot$. Note that we allow paths to configurations $(a_\bot,\ell)$, $\ell \in \mathit{Err}$, since $a_\bot$ represents the empty set of concrete states, and thus does not stand for a concretely reachable error. 

The abstract transition system can be used for checking properties of the concrete program whenever the abstract transformers are safe approximations.

\begin{theorem} \label{th:sound-abstraction} 
Let $P$ be a CFA, $T$ its concrete and $T^\#$ its abstract transition system according to some abstract domain and safe abstract next transformer. Then the following holds: If $T^\#$ is free of errors, so is $T$. 
\end{theorem}



\section{Transformer Constraints} \label{sec:constraints}

The framework of abstract interpretation is used to verify that a program is free of errors. To this end, the abstract transition system is build and inspected for reachable error locations. However, the construction of the abstract transition system and thus the soundness of verification relies on the fact that the abstract transformer safely approximates the concrete transformer, and this in particular means that we verify properties of a program execution using the concrete transformers for next state computation. This assumption is not true anymore when we run our programs on approximate hardware. 

For  a setting with approximate hardware, we have {\em approximate next transformers} $next_{stm}^{AC}$ for (some or all) of our statements. The key question is now the following: Under which conditions on these approximate transformers will our verification result carry over to the AC setting? 
To this end, we need to find out what "properties" of a statement the verification run has actually used. This can be seen in the abstract transition system by looking at the transitions labelled with a specific statement, and extracting the abstract states before and after this statement. A {\em tolerance constraint} for a statement includes all such pairs of abstract states, specifying a number of pre- and postconditions for the statement. 

\begin{definition} Let $T^\# = (Q,q_0,\transo)$ be an abstract transition system of a program $P$, $stm \in Stm$ a statement.  Let $((a_1^i,\ell_1^i) \trans {stm} (a_2^i,\ell_2^i))_{i\in I}$ be the family of transitions in $\trans {stm} \cap (Reach^\# \times Reach^\#)$. 

The {\em tolerance constraint for $stm$ in $T^\#$} is the family of pairs of abstract states $((a_1^i,a_2^i))_{i\in I}$. 
\end{definition}

\noindent While the concrete transformers by safe approximation fulfill all these constraints, the approximate transformers might or might not adhere to the constraints. 

\begin{definition}
A next transformer $next^{AC}_{stm}: \Sigma \rightarrow \Sigma$ {\em fulfills a tolerance constraint} $((a_1^i,a_2^i))_{i\in I}$  if the following property holds for all $i\in I$: 
\[ s \in \gamma(a_1^i) \Rightarrow next^{AC}_{stm}(s) \in \gamma(a_2^i) \ . \]
\end{definition} 

\noindent When programs are run on approximate hardware, the execution will use some approximate and some precise next transformers depending on the actual hardware. For instance, the execution might employ an approximate adder, and thus all statements using addition will be approximate. We let $T^{AC}(P)$ be the transition system of program $P$ constructed by using $next^{AC}_{stm}$ for the approximate statements and standard concrete transformers for the rest.  This lets us now formulate our main theorem about the validity of verification results on AC hardware. 

\begin{theorem} \label{th:sound}
  Let $P$ be a program and $next_{stm}^{AC}$ be a next transformer for $stm$ fulfilling the tolerance constraint on $stm$ derived from an abstract transition system $T^\#(P)$ wrt.\ some abstract domain $Abs$ and safe abstract next transfomers.   Then we get: 

 If $T^\#(P)$ is free of errors, so is $T^{AC}(P)$. 
\end{theorem} 

\noindent {\bf Proof:} Let $((a_1^i,a_2^i))_{i\in I}$ be the tolerance constraint for $stm$ in $T^\#(P)$. Assume the contrary, i.e., there is a path to an error location in $T^{AC}$: $(s_0,\ell_0) \trans {stm_0} (s_1, \ell_1) \trans {stm_1} \ldots \trans {stm_{n-1}} (s_n,\ell_n)$ such that $\ell_n \in Err$. We show by induction that there exists a path $(a_0,\ell_0) \trans {stm_0} (a_1, \ell_1) \trans {stm_1} \ldots \trans {stm_{n-1}} (a_n,\ell_n)$ in the abstract transition system $T^\#$  such that $s_j\in \gamma(a_j)$. 
\begin{description}
\item[Induction base.] $s_0 \in \gamma(a_0)$ since $a_0 = \alpha(\{s_0\})$ and $\{s_0\} \subseteq \gamma(\alpha(\{s_0\}))$ by Galois connection properties. 
\item[Induction step.] Let $s_j\in \gamma(a_j)$, $next_{stm_j}(s_j) = s_{j+1}$ and $(\ell_j,stm_j,\ell_{j+1}) \in E$. Let $S_j = \gamma(a_j)$ (hence $s_j \in S_j$). Now we need to consider two cases: 

Case (1): $stm_j \neq stm$: Then the next transformer applied to reach the next configuration is the standard transfomer.  Thus let $next_{stm_j}(\gamma(a_j)) = S_{j+1}$. By safe approximation of $next^\#$, we get 
$\alpha(S_{j+1}) \sqsubseteq_{Abs} next^\#_{stm_j}(a_j) = a_{j+1}$. 
By monotonicity of $\gamma$: $\gamma(\alpha(S_{j+1})) \subseteq \gamma(a_{j+1})$. 
By Galois connection: $S_{j+1} \subseteq \gamma(\alpha(S_{j+1}))$. 
Hence, by transitivity $s_{j+1} \in \gamma(a_{j+1}) $. 

Case (2): $stm_j = stm$: Let $next^\#_{stm}(a_j) = a_{j+1}$. By definition of tolerance constraint extraction, the pair $(a_j,a_{j+1})$ has to be in the family of tolerance constraints, i.e., $\exists i \in I: (a_j,a_{j+1}) = (a_1^i,a_2^i)$. Since $next^{AC}_{stm}$ fulfills the constraint, $next^{AC}_{stm_j}(s_j) \in \gamma(a_{j+1})$.  \hfill $\Box$
\end{description} 


%
%

\section{Preserving termination}\label{sec:termination}

So far, we have been interested in the preservation of already proven safety properties on approximate hardware. 
Another important issue is the preservation of {\em termination}: whenever we have managed to show that a program terminates on precise hardware, we would also like to get constraints that guarantee termination on AC hardware. In order to extend our approach to termination, we make use of an approach for encoding termination proofs as safety properties \cite{DBLP:conf/pldi/CookPR06}. 

We start with explaining standard termination proofs. Nontermination arises when we have loops in programs and the loop condition never gets false in a program execution. In control flow automata, a {\em loop} is a sequence of locations $\ell_0, \ldots, \ell_n$ such that there are statements $stm_i$, $i=0..n-1$, with $\ell_i \trans {stm_i} \ell_{i+1}$ and $\ell_0 = \ell_n$. In this, a location $\ell = \ell_i$ is said to be on the loop. Every well-structured loop has a condition and a loop body: the {\em start} of a loop body is a location $\ell$ such that there are locations $ \ell', \ell''$ and a boolean condition $b$ s.t.\ $\ell' \trans {assume\ b} \ell$ and $\ell' \trans {assume\ !b} \ell''$ are in the CFA and $\ell'$ is on a loop, but either $\ell''$ is not on a loop, or is on a different loop. 
Basically, we just consider CFAs of programs constructed with while or for constructs, not with gotos or recursion. However, the latter is also possible when the verification technique used for proving termination covers such programs. 

\begin{definition}
A {\em non-terminating run} of a CFA $P=(L, \ell_0, E, \mathit{Err})$ is  an infinite sequence of configurations and statements $(s_0,\ell_0) \trans{stm_1} (s_1,\ell_1) \trans {stm_2} \ldots$ in the transition system $T(P)$. If $P$ has no non-terminating runs, then $P$ {\em terminates}. 
\end{definition} 

\begin{proposition}
In every non-terminating run, at least one loop start $\ell$ occurs infinitely often.
\end{proposition}

\noindent We assume some standard technique to be employed for proving termination. Such techniques typically consist of (a) the synthesis of a termination argument, and (b) the check of validity of this termination argument. Termination arguments are either given as monolithic ranking functions or as disjunctively well-founded transition invariants \cite{DBLP:conf/lics/PodelskiR04}. Here, we will describe the technique for monolithic ranking functions. 

\begin{enumerate}
	\item For every loop starting in $\ell$, define a {\em ranking function} $f_\ell$ on the program variables, i.e., $f_\ell: \Sigma \rightarrow W$, where $(W,\leq)$ is a well-founded order with least element $\bot_W$. 
	\item Show $f_\ell$ to decrease with every loop execution, i.e., if $(s,\ell) \trans {stm_1} (s_1,\ell_1) \trans {stm_2} \ldots \trans {stm_n} (s',\ell)$ is a path in $T(P)$, show $f_\ell(s') < f_\ell(s)$. 
	\item Show $f_\ell$ to be greater or equal than the least element of $W$ at loop start, i.e., for all starts of loop bodies $\ell$ and $(s,\ell) \in Q_{T(P)}$, show $f_\ell(s) \geq \bot_W$. 
\end{enumerate}

\noindent If properties (2) and (3) hold, we say that the ranking function is {\em valid}. Note that we are not interested in computing ranking functions here; we just want to make use of existing verification techniques. The following proposition states a standard result for ranking functions (see e.g.\ \cite{Manna1996,Apt2009}). 

\begin{proposition} Let $P$ be a program. If every loop $\ell$ of $P$ has a valid ranking function, then $P$ terminates.
\end{proposition} 

\noindent As an example consider the program $Sum$ on the left of Figure \ref{fig:instr}. It computes the sum of all numbers from 0 up to some constant $N$. It terminates since variable $i$ is constantly increased. As ranking function we can take $N-i$ using the well-founded ordering $\mathbb{N}$. 

In order to encode the above technique in terms of assertions, we instrument a program $P$ along the lines used in the tool \textsc{Terminator} \cite{DBLP:conf/pldi/CookPR06} thereby getting a program $\widehat{P}$ as follows. Let $\mathit{Var} = \{x_1, \ldots, x_n\}$ be the set of variables occuring in the program. At starts of loop bodies $\ell$ we insert
\begin{verbatim}
  if (!(f_l(x1, ..., xn) >= bot_W)
     ERR:
  old_x1 := x1;
  ...
  old_xn := xn;
\end{verbatim} 
\noindent and at loop ends we insert 
\begin{verbatim}
  if (!(f_l(x1, ..., xn) <_W f_l(old_x1, ..., old_xn))
     ERR:
\end{verbatim} 
\noindent when given a ranking function $f_\ell$ and a well-founded ordering $(W,<_W)$ with bottom element \verb+bot_W+.  

\begin{proposition}
  If $\widehat{P}$ if free of errors, then $P$ terminates.
\end{proposition}

\noindent Hence we can use standard safety proving for termination as well (once we have a ranking function), and thereby derive tolerance constraints.  In the left of Figure \ref{fig:instr} we see the instrumented version of program $Sum$. Here, we have already applied an optimization: we only make a copy of variable \verb+i+ since the ranking function only refers to \verb+i+ and \verb+N+, and \verb+N+ does not change anyway. 

\begin{figure}[t]
\begin{minipage}[t]{.1\textwidth}\ 
\end{minipage}
\begin{minipage}[t]{.4\textwidth}
\begin{verbatim}
sum=0;
i=0;
while (i<N) { 



    sum=sum+i;
    i=i+1; 
    
    
}
\end{verbatim}
\end{minipage} \quad 
\begin{minipage}[t]{.4\textwidth}
\begin{Verbatim}[commandchars=\\\[\]]
sum=0;
i=0;
while (i<N) { 
   \color[red][if (!(N-i > 0))]
      \color[red][ERR: ;]
   \color[red][old_i=i;] 
   sum=sum+i;
   i=i+1; 
   \color[red][if (!(N-i < N-old_i))]
      \color[red][ERR: ;]
}
\end{Verbatim}
\end{minipage}
\caption{Program $Sum$ (on the left)  and its instrumented version (on the right).}
\label{fig:instr}
\end{figure}

\section{Constraint Extraction for Predicate Analysis}

Section \ref{sec:constraints} has formally defined the extraction of tolerance constraints from abstract transition systems and has proven its soundness. Now, we will take a closer look at constraint extraction in practice. To this end, we choose an instance of the abstract interpretation framework, namely predicate abstraction \cite{Graf1997,DBLP:journals/sttt/BallPR03}. Furthermore, instead of deriving constraints for statements, we derive constraints for {\em operators} since in practice we do not have specific hardware for whole statements but just for the operations used in expressions within a statement. 

We start with defining predicate abstraction. 
For this, we fix a set of predicates ${\cal P}$ over $\mathit{Var, Cmp}, \mathbb{Z}$ and $\mathit{Ops}$. In practice, these predicates will be incrementally computed by a counter-example-guided abstraction refinement approach \cite{DBLP:conf/popl/HenzingerJMM04} which we just assume to exist (and which is provided by the tool that we employ for our experiments).  We define $\neg {\cal P} := \{ \neg p \mid p \in {\cal P}\}$ and let the abstract domain $Abs$ be conjunctions of predicates or their negations (also directly written as set of literals, hence $\emptyset$ is true, ${\cal P} \cup \neg {\cal P}$ is false):   

\[ (\{ \bigwedge _{q\in Q} q \mid Q \subseteq {\cal P} \cup \neg {\cal P} \}, \Rightarrow) \]

\noindent The Galois connection is given by letting $\alpha(S) = \{q \in {\cal P} \cup \neg {\cal P} \mid \forall s \in S: s \models q \}$ and $\gamma(Q) := \{s \in \Sigma \mid \forall q \in Q: s\models q \}$. We write $s\models Q$ iff $s\models q$ for all $q\in Q$. For the definition of the abstract next transformers see for instance \cite{DBLP:journals/sttt/BallPR03}. Note that tolerance constraints in this domain take the form $(Q_1,Q_2), Q_j \subseteq {\cal P} \cup \neg {\cal P}, j \in \{1,2\}$. 

\tikzstyle{leer} = [rectangle, text width=1em, text centered]
\tikzstyle{block} = [rectangle, draw, node distance=1cm, 
    text width=7em, text centered, minimum height=2.3em]
\tikzstyle{line} = [draw, -latex']
 \tikzstyle{transition}=[rectangle,thick,draw,-latex']
\tikzstyle{circ} = [circle, thick,draw=blue!75,fill=blue!20,minimum size=2mm]

\begin{figure}[t]
\begin{tikzpicture}[node distance = 0.8cm, auto]
\tikzstyle{every state}=[draw=blue!75,fill=blue!20,minimum size=4mm]
\node[leer] (tsu) {};
\node[leer] (foo) [right=of tsu] {};
\node[state](0) [right=of foo, label=below:\textcolor{purple}{$true$}] {$\ell_0$};
\node[leer] (blub) [right=of 0] {};
\node[state](1)[right=of blub, label=right:\textcolor{purple}{$true$}]{$\ell_1$};
\node[state](2)[below=of 1, label=left:\textcolor{purple}{$j\geq 0$}]{$\ell_2$};
\node[state](3)[below=of 2, label=right: \textcolor{purple}{$ j\geq 0 \wedge j \leq 989$}]{$\ell_3$};
\node[state](4)[below=of 3, label=right: \textcolor{purple}{$j \geq 0 \wedge j \leq 999$}]{$\ell_4$};
\node[leer] (bla) [below=of 4] {};
\node[state](5)[left=of bla, label=right: \textcolor{purple}{$j\geq 0$}]{$\ell_5$};
\node[state](err)[right=of bla, label=right:\textcolor{purple}{$false$}]{$err$};
\node[leer] (blub) [right=of 2] {};
\node[state](6)[right=of blub, label=right: \textcolor{purple}{$j \geq 0$}]{$\ell_6$};
\path [transition] (0) edge node[above]{int arr[1000];} (1);
\path [transition] (2) edge node[above]{$\neg$(j$<$990)} (6);
\path [transition] (1) edge node[right]{int j=0;} (2);
\path [transition] (2) edge node[right]{j$<$990} (3);
\path [transition] (3) edge node[right]{j:=j+10;} (4);
\path[transition] (4) edge node[left]{j $\geq$ 0 $\wedge$ j $<$ 1000} (5);
\path[transition] (4) edge node[right]{$\neg$(j $\geq$ 0 $\wedge$ j $<$ 1000)} (err);
\path[transition] (5) edge[bend left] node[left] {arr[j]:=0;} (2);
\end{tikzpicture}
\caption{Abstract transition system of program $Array$}
\label{fig:art}
\end{figure}
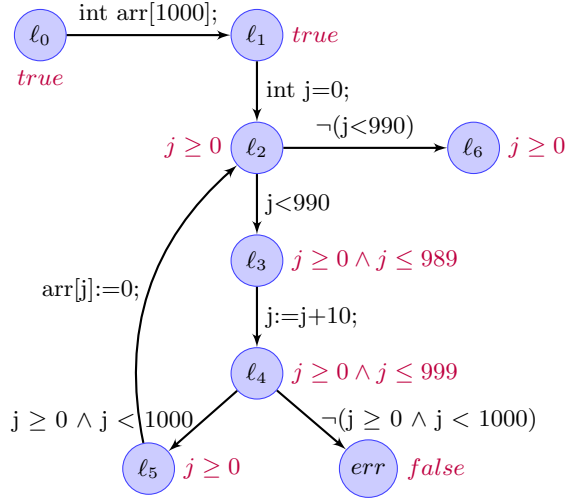

This abstract domain can be used to show program $Array$ from Figure \ref{fig:array} to be free of errors. Figure \ref{fig:art} shows the abstract transition system of program $Array$ using the predicate set ${\cal P} = \{j \geq0, j \leq 989, j \leq 999\}$. The predicates holding in an abstract configuration $(a, \ell)$, i.e., the abstract state $a$, are written next to the purple location. We see that the location labeled $err$ occurs in the graph, but the abstract state in this configuration is $a_\bot = false$, and, thus, we say that this error is not reachable.

%

 For the extraction of tolerance constraints for operators $op \in Op$, we assume our statements to take the form of {\em three-address code} (3AC) \cite{DBLP:books/aw/AhoSU86}. In three-address code form, all operators $op$ occur in programs only in statements $v:= a \mathop{op} b$, where $a$ and $b$ are variables or constants. Every program can be brought in such a form (e.g., intermediate representations generated during compilation take this form). We use this 3AC form  because we need to isolate operators, and only have statements with one (possibly approximate) operator in. Note that program $Array$ is in 3AC form. 

Furthermore, the tolerance constraints, i.e., pre- and postcondition predicates, derived from abstract transition systems are specified over the {\em program variables}. As an example, take the operator $+$.  In the program $Array$ this operator occurs in the statement \verb#j:=j+10#. The tolerance constraint for this statement derived from the abstract transition system in Figure \ref{fig:art} is $(j \geq 0 \wedge j \leq 989, j \geq 0 \wedge j \leq 999)$. This constraint refers to the program variable $j$. If the approximate adder used for $+$ has inputs $x$ and $y$ and output $z$, this constraint first of all needs to be brought into a form using variables $x$, $y$ and $z$. This is achieved using the following replacement operator. 

\begin{definition}
Let \(Q \in {\cal P} \cup \neg {\cal P}, p \in Q, v_1,v_2\in Var\).
The predicate \(p[v_2\symTC v_1]\) is obtained from $p$ by replacing  all occurrences of \(v_2\) by \(v_1\). We lift this to sets by letting \(Q[v_2\symTC v_1]:=\{q[v_2\symTC v_1]\mid q\in Q\}\).
For constants \(c\in\mathbb{Z}\), we define \(Q[c \symTC v_1]:=Q\cup\{v_1 = c\}\).
\end{definition}

\begin{proposition} For all $q\in {\cal P} \cup \neg {\cal P}$ such that $x\notin vars(q)$: 
\[ s[x:=s(u)] \models q[u \symTC x] \quad \Leftrightarrow \quad s \models q \]
\end{proposition} 

\noindent For constraint $(Q_1, Q_2) = (j \geq 0 \wedge j \leq 989, j \geq 0 \wedge j \leq 999)$, statement \verb#j:=j+10# and adder with inputs $x$ and $y$, output $z$, the replacement we need to make is $(Q_1[j\symTC x, 10 \symTC y], Q_2[j \symTC z]) = (x \geq 0 \wedge x \leq 989 \wedge y=10, z \geq 0 \wedge z \leq 999)$. This is the constraint which ultimately needs to be checked for the approximate hardware. 
In the following we assume all binary operators to have signature $(x : \mathbb{Z}, y: \mathbb{Z}) \rightarrow (z: \mathbb{Z})$, $x, y$ and $z$ to not occur as variables in the program nor in the predicates and use $ (\widehat{Q}_1, \widehat{Q}_2)$ to refer to the constraints obtained after the replacement.

\begin{definition} 
An approximate operator $op^{AC}: \mathbb{Z} \times \mathbb{Z} \rightarrow \mathbb{Z}$ adheres to a tolerance constraint $(\widehat{Q}_1, \widehat{Q}_2)$ (i.e., over $x,y$ and $z$) if 
\[ \forall s \in \Sigma: s\models \widehat{Q}_1 \Rightarrow s[z:=op^{AC}(s(x),s(y))] \models \widehat{Q}_2 \]
\end{definition} 

\noindent Adherence to constraints by operators implies adherence to constraints by statements using these operators. 

\begin{lemma} \label{lem:adhere} Let $(Q_1,Q_2)$ be a tolerance constraint extracted from $T^\#$ for $stm \equiv u:= v \mathop{op} w$. If $op^{AC}$ adheres to $(Q_1[v\symTC x, w \symTC y],Q_2[u \symTC z])$, then $next_{stm}^{AC} :\equiv u:= v \mathop{op^{AC}} w$ adheres to $(Q_1,Q_2)$.
\end{lemma}

\noindent {\bf Proof}: We need to show that $next_{stm}^{AC}$ adheres to $(Q_1,Q_2)$. We first of all take the definition of it and rewrite it a little.
\begin{eqnarray*}
s \in \gamma(Q_1) & \Rightarrow & next^{AC}_{stm}(s) \in \gamma(Q_2) \\
                           & \Leftrightarrow &   \quad \{ \mbox{ definition of } next_{stm}^{AC}  \quad \} \\
s \in \gamma(Q_1) & \Rightarrow & s[u:=op^{AC}(s(v),s(w))] \in \gamma(Q_2) \\
                          & \Leftrightarrow & \quad \{ \mbox{ definition of } \gamma \quad \} \\
s \models Q_1 & \Rightarrow & s[u:=op^{AC}(s(v),s(w))] \models Q_2
\end{eqnarray*}
The last implication is now shown as follows:
\begin{eqnarray*}
s \models Q_1 & \Rightarrow & s[x:=s(v),y:=s(w)] \models Q_1[v\symTC x, w\symTC y] \\
                    & \Rightarrow & s[x:=s(v),y:=s(w),z:=op^{AC}(s(v),s(w))] \models Q_2[u\symTC z] \\
                    & \Rightarrow & s[x:=s(v),y:=s(w),u:=op^{AC}(s(v),s(w))] \models Q_2 \\
                    & \Rightarrow & s[u:=op^{AC}(s(v),s(w))] \models Q_2 
\end{eqnarray*}
\hfill $\Box$ 

\noindent This finally gives us our main soundness result for predicate analysis which is an immediately corollary of Lemma \ref{lem:adhere} and Theorem \ref{th:sound}. 

\begin{corollary} Let $T^\#$ be an abstract transition system constructed using safe approximations and let all approximate operators $op^{AC}$ adhere to the constraints derived from $T^\#$. Then: 
If $T^\#$ is free of errors, so is $T^{AC}$. 
\end{corollary}

\paragraph{Implementation.} 
As proof of concept we integrated our proposed constraint extraction into the software analysis tool \caps{CPAchecker}~\cite{CPAchecker2011}, a tool for C program analysis which is configurable to abstract interpretation based analyses.
Mainly, we added a constraint extraction algorithm   plus some additional helper classes.
Our constraint extraction algorithm builds on top of \caps{CPAchecker}'s predicate analysis which uses the technique of adjustable block enconding \cite{ABE}, a technique which allows to specify at which locations an abstraction should be computed. 
For our extraction we need to make sure that we have an abstract state immediately before and after each statement which uses the operation of interest \(op\)\footnote{The operation of interest is made configurable in \caps{CPAchecker}.}.
To identify these abstraction points and later the tolerance constraints, we first need to identify the statements using the operation \(op\).
Afterwards, we run \caps{CPAchecker}'s standard predicate analysis which provides us with an abstract reachability graph (ARG), a structure similar to the abstract transition system. In the ARG, the predicates are given in SMT-Lib format \cite{SMTLIB} since \caps{CPAchecker} is using state-of-art SMT solvers for predicate analysis. 
From the ARG, we extract the tolerance constraints and write one SMT file per constraint \((Q_1, Q_2)\) which is in the input format required by our next tool building the hardware checker.
The SMT file mainly contains the description of \((Q_1, Q_2)\) pairs plus additional information about the signature of the statement for which the constraint was extracted.
The signature is needed by the next tool to construct \((\widehat{Q_1},\widehat{Q_2})\).
%

To run the tolerance constraint extraction within \caps{CPAchecker}, one can use the configuration file \texttt{predicateAnalysis-ToleranceConstraintsExtraction-} \\ \texttt{PLUS.properties} that we used in our evaluation to extract tolerance constraints for additions.

\section{Constraint Checking}

\begin{figure}
	\centering
	\includegraphics[width=0.8\textwidth]{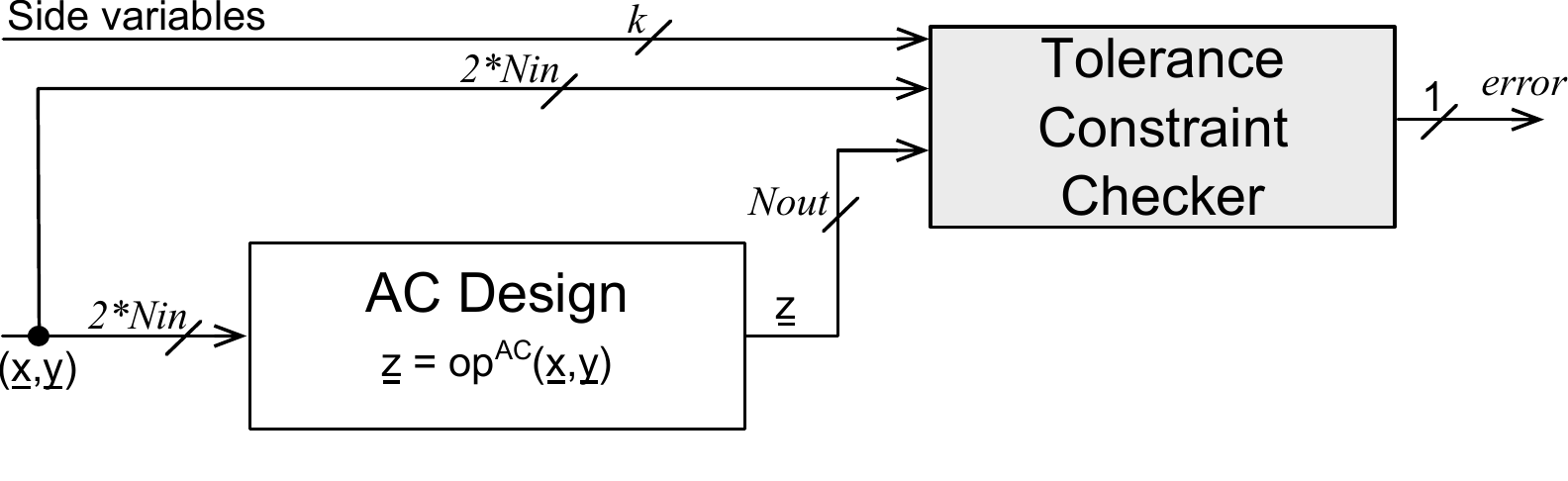}
	\caption{Adherence Checker combining AC design with Tolerance Constraint Checker}
	\label{fig:Checker}
\end{figure}

The final step of our technique is the check of the extracted constraints on actual hardware designs of approximate operations.  For simplicity of representation, we restrict the following explanations to the case of a single constraint\footnote{A generalization to a family of constraints is straightforward.}. The input to the checking phase thus consists of a constraint \((Q_1, Q_2)\), an approximate operator $op^{AC}$ and the corresponding program statement $u:= v \mathop{op} w$. The checking of the constraint on a given hardware design with inputs $x,y$ and output $z$ (in our case specified in Verilog) proceeds in three steps:
\begin{description}
\item[Mapping] The mapped tolerance constraint \((\widehat{Q}_1, \widehat{Q}_2) = (Q_1[v\symTC x, w\symTC y], Q_2[u\symTC z])\) is constructed.  As a result, the tolerance constraint \((\widehat{Q}_1, \widehat{Q}_2)\) uses the variables $x$, $y$ and $z$ when referring to the inputs and output of $op^{AC}$. Additional variables of the program (besides $u, v$ and $w$) may still occur in the constraint which are not used in the hardware design. We denote these variables as {\em side variables}.
\item[Transformation] The mapped constraint is transformed into Verilog code giving a {\em checker circuit}. The checker circuit is created as Verilog code in two steps.
First, the logical formulae of the tolerance constraints  are compiled to Verilog code (see \cite{PauckBachelorarbeit}). In this, side variables are treated like other inputs. We then fix a single output of the checker called $error$   by setting 
$error := \neg (\widehat{Q}_1 \Rightarrow \widehat{Q}_2)$. 
\item[Combination] The generated tolerance constraint checker is afterwards combined with the hardware design of $op^{AC}$ into an {\em adherence checker}. For our examples, the AC hardware designs are also given in Verilog.
The combination is done using a top module that contains and wires the design of $op^{AC}$ and the tolerance checker as sub-modules.
The wiring is done as depicted in Figure \ref{fig:Checker}.
\end{description}

\noindent The resulting circuit is afterwards checked for safety, i.e., that for no combinations of values on the primary inputs the error flag is raised.
This step can be done using standard hardware verification techniques (unsatisfiability checking).

As an example, consider again program $Array$ given on the left side of Figure \ref{fig:array}. The tolerance constraint extracted for operator $+$ is $(j \geq 0 \wedge j \leq 989, j\geq 0 \wedge j \leq 999)$ and the program statement is \verb#j:=j+10#. In SMT-Lib format, the constraint is  
\begin{verbatim}
		(define-fun Q_1 () Bool (and (<= 0 |main::j|) (<= |main::j| 989)))
		(define-fun Q_2 () Bool (and (<= 0 |main::j@1|) 
		                                            (<= |main::j@1| 999)))
\end{verbatim}
The structural mapping of the variables is represented as $[|main::j|\symTC x]$, $[10 \symTC y]$ and $[|main::j@1| \symTC z]$.
	As a result, the mapped constraint can be represented as follows.
\begin{verbatim}
		(define-fun mappedQ_1 () Bool (and (and (<= 0 x) (<= x 989)) 
		                                        (= y 10)))
		(define-fun mappedQ_2 () Bool (and (<= 0 z) (<= z 999)))
\end{verbatim}
Figure \ref{fig:checker} gives the Verilog code of the checker circuit belonging to this mapped constraint. Note that the length of the input vectors have to be adapted to fit the one provided by the hardware design of $op^{AC}$.

\begin{figure}[t]
\begin{minipage}[t]{.6\textwidth}
\begin{verbatim}
module TCChecker(x,y,z, error);
  parameter Nin = 32;
  parameter Nout = 33;

  input [Nin-1:0] x;
  input [Nin-1:0] y;
  input [Nout-1:0] z;
  output error;

  wire term__1;
  wire term__2;
  wire term__3;
  wire term__4;
  wire Q__1;
  wire Q__2;
  wire pre__gen_0;

  assign term__1 = (x <= 989);
  assign term__2 = (0 <= x);
  assign Q__1 = (term__2 && term__1);
  assign term__3 = (z <= 999);
  assign term__4 = (0 <= z);
  assign Q__2 = (term__4 && term__3);
  assign pre__gen_0 = (y == 10);

  assign error = !(( !(Q__1 && pre__gen_0) || Q__2));
endmodule
\end{verbatim}
\end{minipage}\quad 
\begin{minipage}[t]{.25\textwidth}
\begin{verbatim}
module AdherenceChecker(
  input [31:0] inp1,
  input [31:0] inp2,
  output errorBit);

  wire[32:0] outp;

  Adder add(
    .x(inp1),
    .y(inp2),
    .z(outp)
  );

  TCChecker check(
    .x(inp1),
    .x(inp2),
    .z(outp),
    .error(errorBit)
  );
endmodule
\end{verbatim}
\end{minipage}
\caption{Verilog code of checker circuit (left) and its combination with Adder (right)}
\label{fig:checker}
\end{figure}

\paragraph{Experiments.}

In our experiments, we used the software analysis tool \caps{CPAchecker} to extract the tolerance constraints from a verification run.
We employed the tools {\em Yosys} \cite{Yosys} and {\em ABC} \cite{berkeley2005system} for synthesis and generation of a CNF formula that encodes the value of the error flag in dependence on all the inputs.
Using PicoSAT \cite{biere2013picosat}, we checked the unsatisfiability of the formula, denoting that the error flag is never raised, i.e., the tolerance constraints are met by the implementation.

In the following, we give the results of our experiments.
In our experiments we studied tolerance constraints for addition (since this is the only operation for which approximate hardware is currently publicly available).
While it is often accepted that in approximate computing a computation result is not functional equivalent with a precise computation result, a approximate computation must still well-behave.
For example, memory accesses should remain safe or it should still terminate or stick to a certain protocol.
That is why, during program verification we considered one of these properties instead of functional behavior.

We extracted tolerance constraints from the verification of a number of handcrafted programs (including our three examples) and some programs from the subcategory \verb+ControlFlow+ and \verb+ProductLines+ of the SV-COMP\footnote{Some additions first had to be brought in three-address code form and in some programs we replaced some constant assignments by proper addition.} \cite{SVComp2015}.
We chose our programs to get tolerance constraints from a variety of verification problems and are very well aware that these programs are no typical candidates for approximate computing.
The handcrafted programs \texttt{AddOne}, \texttt{EvenSum}, and \texttt{MonotonicAdd} should examine the addition of positive numbers. 
Programs \texttt{sum}, \texttt{quotient}, and \texttt{mirror\_matrix} use the previously described technique to encode termination proofs with assertions.
To artificially enforce a difference between the behavior of the approximate adder, we used program \texttt{SpecificAdd} which checks that the addition of \(30+50\) is indeed \(80\).
The programs from the SV-COMP (the last 10 programs shown in Table~\ref{tab:eval}) check protocol properties, e.g. correct locking behavior.

We checked the tolerance constraints on a standard, non-approximate ripple carry adder (RCA) and a set of approximate adders provided by the Karlsruhe library of \cite{ShafiqueAHH15} (called ACA-I \cite{verma2008variable}, ACA-II (ACA\_II\_N16\_Q4)\cite{kahng2012accuracy}, ETAII \cite{zhu2009enhanced}, GDA \cite{ye2013reconfiguration} and GeAr).
Table \ref{tab:eval} shows our results.
For each program, we show the number of additions \(\#+\), the number of program statements \(\#stm\), the number of constraints extracted \(\#tc\) and whether an adder meets the tolerance constraints \(\checkmark\) or not \(\times\).

\begin{table}[htbp]
  \caption{Results of experiments}
	\label{tab:eval}
	\centering
		\begin{tabular} {|l | r| r |r |c |c |c |c| c| c|}
		\hline
		program & \#+ & \#stm & \#tc & RCA & ACA-I & ACA-II & ETAII & GDA & GeAr\\
		\hline
		Array & 1 & 15 & 1 & \(\checkmark\) & \(\checkmark\) & \(\checkmark\) & \(\checkmark\) & \(\checkmark\) & \(\checkmark\)\\
		AddOne & 1 & 22 & 1 & \(\checkmark\) & \(\times\) & \(\times\) & \(\times\) & \(\times\) & \(\times\)\\
		Attach/Detach & 1 & 26 & 1 & \(\checkmark\) & \(\checkmark\) & \(\checkmark\) & \(\checkmark\) & \(\checkmark\) & \(\checkmark\)\\
		EvenSum & 2 & 24 & 4 & \(\checkmark\) & \(\times\) & \(\times\) & \(\times\) & \(\times\) & \(\times\)\\
		MonotonicAdd & 1 & 20 & 1 & \(\checkmark\) & \(\times\) & \(\times\) & \(\times\) & \(\times\) & \(\times\)\\
		SpecificAdd & 1 & 13 & 1 & \(\checkmark\) & \(\checkmark\) & \(\times\) & \(\checkmark\) & \(\checkmark\) & \(\checkmark\)\\
		\hline
		sum & 2 & 26 & 2 & \(\checkmark\) & \(\times\) & \(\times\) & \(\times\) & \(\times\) & \(\times\)\\
		quotient & 2 & 35 & 2 & \(\checkmark\) & \(\times\) & \(\times\) & \(\times\) & \(\times\) & \(\times\)\\
		mirror\_matrix & 2 & 42 & 2 & \(\checkmark\) & \(\times\) & \(\times\) & \(\times\) & \(\times\) & \(\times\)\\
		\hline
		locks\_5 & 5 & 114 & 31 & \(\checkmark\) & \(\checkmark\) & \(\checkmark\) & \(\checkmark\) & \(\checkmark\) & \(\checkmark\)\\
		locks\_8 & 8 & 171 & 255 & \(\checkmark\) & \(\checkmark\) & \(\checkmark\) & \(\checkmark\) & \(\checkmark\) & \(\checkmark\)\\
		\hline
		cdaudio & 13 & 1888 & 23 & \(\checkmark\) & \(\checkmark\) & \(\checkmark\) & \(\checkmark\) & \(\checkmark\) & \(\checkmark\)\\
		diskperf & 19 & 981 & 12 & \(\checkmark\) & \(\checkmark\) & \(\checkmark\) & \(\checkmark\) & \(\checkmark\) & \(\checkmark\)\\
		floppy4 & 31 & 1370 & 34 & \(\checkmark\) & \(\checkmark\) & \(\checkmark\) & \(\checkmark\) & \(\checkmark\) & \(\checkmark\)\\
		kbfiltr2 & 11 & 759 & 15 & \(\checkmark\) & \(\checkmark\) & \(\checkmark\) & \(\checkmark\) & \(\checkmark\) & \(\checkmark\)\\
		\hline
		minepump\_s5\_p64 & 2 & 741 & 3 & \(\checkmark\) & \(\checkmark\) & \(\checkmark\) & \(\checkmark\) & \(\checkmark\) & \(\checkmark\)\\
		minepump\_s5\_simulator & 2 & 811 & 3 & \(\checkmark\) & \(\checkmark\) & \(\checkmark\) & \(\checkmark\) & \(\checkmark\) & \(\checkmark\)\\
		\hline
		clnt\_4 & 13 & 575 & 18 & \(\checkmark\) & \(\checkmark\) & \(\checkmark\) & \(\checkmark\) & \(\checkmark\) & \(\checkmark\)\\
		srvr\_8 & 19 & 668 & 14 & \(\checkmark\) & \(\checkmark\) & \(\checkmark\) & \(\checkmark\) & \(\checkmark\) & \(\checkmark\)\\
		\hline
		\end{tabular}

\end{table}

Our first observation is that except for program \texttt{SpecificAdd} which we created to show a different behavior between the approximate adders either or all approximate adders meet the extracted tolerance constraint or none of them.
This is because all approximate adders use the same principle: reduction of the carry chain.
In their addition, they use a set of subadders and the carry bit of the previous subadder is either dropped or imprecisely predicted.
The effect of this reduction only shows off for specific numbers and these specific numbers differ among the approximate adder.
Hence, adding 30 and 50  failed only in the approximate adders ACA-II.

Interestingly, the approximate adders meet the extracted tolerance constraints for all of the SV-COMP programs. 
On the one hand, not all additions in the programs have an effect on the correctness of the program (and thus verification imposes no constraints on them).
On the other hand, typically the additions considered during verification which had an effect increase a variable value in the range \([0,9]\) by one which can be computed precisely by the first subadder of all approximate adders.

For our own programs, one can see that all sorts of cases occur: all approximate adders satisfy the extracted constraints (as is the case for program $Array$), some do and some do not (on program {\em SpecificAdd}), and all do not. An instance of the latter case is our example program $AddOne$ from the right of Figure \ref{fig:array}. 
The variable $u$ which is increased by 1 can be any positive integer (it is an input). The derived constraint for operator $+$ is \(((1\leq u)[u\symTC x, 1\symTC y], (sum\neq0)[sum \symTC z])=(1\leq x \wedge y=1, z\neq0)\). 
For our verification of the property, we require that the increase of that variable does not result in value zero, which can be the case if the carry propagation is imprecise. Thus, here the approximate designs fail to satisfy the constraint. Hence, an execution of the program on approximate hardware with these adders could reach the error state. 
The imprecise carry propagation is also the reason why the approximate adders cannot guarantee termination of programs \texttt{sum}, \texttt{quotient}, and \texttt{mirror\_matrix}.
For termination all three programs rely on an addition which is strongly monotonic up to a certain threshold (maximal int value).
However, due to the imprecise carry propagation an addition of two positive integers may result in value zero.

\section{Discussion}\label{sec:discuss}
To compute requirements on AC hardware with the help of program verification, further approaches are conceivable.
For example, one could model the approximate operation as a function call.
This means, the approximate operations in a program, e.g.\ the approximate addition, must be replaced by a call to corresponding function. 
Now, one applies a verification technique, e.g.\ \cite{Sery2011},  which computes function summaries \cite{Hoare1971}.
The function summary for the approximate operation, in principle a description of a pre-/postcondition pair, gives us the constraint on the AC hardware.

In another alternative one would also model the approximate operation as a function call, but now one assumes that the behavior of the function \(\mathrm{approx}\) modeling the approximate operation is unknown.
In this case, one may use a technique like \cite{Albarghouthi2016} which tries to generate the weakest specification for the function \(\mathrm{approx}\) which still ensures program correctness w.r.t. the desired property.
The specification for the function \(\mathrm{approx}\) which is in principle an encoding of a pre-/postcondition pair describes the requirement on the AC hardware.

We are confident that both alternatives could be used with our general approximation tolerance constraints approach.
To use those alternatives the function summary and the inferred specification must be transformed into a tolerance constraint checker.
We think this is feasible because \cite{Sery2011} and \cite{Albarghouthi2016} already seem to use logic formulae to express the function summary and the specification.

For this paper, we decided to use abstract interpretation as a first example to generate the tolerance constraints for AC hardware. 
The disadvantage of abstract interpretation is that we might get multiple constraints. 
The two alternatives only generate one constraint.
In practice, we solved this problem such that multiple constraints are conjuncted into a single constraint during the generation of the tolerance constraint checker. 
On the other hand, we do not need to transform the approximate operations into function calls and the generation of three-address code is rather standard for compilers.
Another reason is that we are already familiar with abstract interpretation.
Additionally, the verification tool \caps{CPAchecker} which we typically use for verification is based on abstract interpretation and analyses functions via in-lining.

\section{Conclusion}

In this paper, we have proposed a new way of making software verification robust against approximate hardware. Its basic principle is the derivation of constraints on AC hardware from verification runs. We have shown our technique to be sound, i.e., shown that the verification result carries over to a setting with AC hardware when the hardware satisfies the derived constraints. First experimental results have shown that the verification result often but not always carries over. More experiments are, however, necessary when further AC implementations of operations -- besides approximate adders -- become available.

\bibliographystyle{splncs03}
\bibliography{literature}

\end{document}